\documentclass[twocolumn,showpacs,amsmath,amssymb,aps,floatfix]{revtex4}

\usepackage{graphicx}% Include figure files
\usepackage{dcolumn}% Align table columns on decimal point
\usepackage{bm}% bold math

\begin{document}

\title{Statistical properties of thermodynamically predicted RNA secondary structures in viral genomes}

%%%%%%% I VERSIONE%%%%%%%
\author{Marco Span\`o$^\dagger$, Fabrizio Lillo$^{\dagger \star}$, Salvatore Miccich\`e$^\dagger$, Rosario N. Mantegna$^\dagger$}                     % Do not remove
\affiliation{$~^\dagger$ Dipartimento di Fisica e Tecnologie Relative, Universit\`a di Palermo, Viale delle Scienze, I-90128, Palermo, Italy \\
                   $~^\star$ Santa Fe Institute, 1399 Hyde Park Road, Santa Fe, NM 87501, USA}

%%%%%%% II VERSIONE%%%%%%%
%\author{Marco Span\`o}                     % Do not remove
%\affiliation{Dipartimento di Fisica e Tecnologie Relative, Universit\`a di Palermo, Viale delle Scienze, I-90128, Palermo, Italy}
%
%\author{Fabrizio Lillo,}                     % Do not remove
%\affiliation{Dipartimento di Fisica e Tecnologie Relative, Universit\`a di Palermo, Viale delle Scienze, I-90128, Palermo, Italy \\
%
%                  Santa Fe Institute, 1399 Hyde Park Road, Santa Fe, NM 87501, USA}
%
%\author{Salvatore Miccich\`e}                     % Do not remove
%\affiliation{Dipartimento di Fisica e Tecnologie Relative, Universit\`a di Palermo, Viale delle Scienze, I-90128, Palermo, Italy}
%
%\author{Rosario N. Mantegna}                     % Do not remove
%\affiliation{Dipartimento di Fisica e Tecnologie Relative, Universit\`a di Palermo, Viale delle Scienze, I-90128, Palermo, Italy}

\date{\today}

\begin{abstract}
By performing a comprehensive study on 1832 segments of 1212 complete genomes of viruses, we show that in viral genomes the hairpin structures of thermodynamically predicted RNA secondary structures are more abundant than expected under a simple random null hypothesis. The detected hairpin structures of RNA secondary structures are present both in coding and in noncoding regions for the four groups of viruses categorized as dsDNA, dsRNA, ssDNA and ssRNA. For all groups hairpin structures of RNA secondary structures are detected more frequently than expected for a random null hypothesis in noncoding rather than in coding regions. However, potential RNA secondary structures are also present in coding regions of dsDNA group. In fact we detect evolutionary conserved RNA secondary structures in conserved coding and noncoding regions of a large set of complete genomes of dsDNA herpesviruses.

\end{abstract}

\pacs{
      {87.18.-h}{Biological complexity} ,
      {87.15.bd}{Secondary structure}   ,
      {87.15.Qt}{Sequence analysis}
     } % end of PACS codes

\maketitle
\section{Introduction}
\label{intro}
In recent years the discovery of the regulatory role of short RNA sequences has changed the view about the biological role of RNA in living organisms \cite{Eddy2001}. For a long time it was assumed that RNA had only an ancillary role in protein synthesis. Today biologists know several regulatory mechanisms fully controlled by RNA short sequences \cite{Mattick2003} often characterized by a typical secondary structure presenting a certain number of hairpin structures. A RNA hairpin structure is a secondary structure where a double stranded region of a single stranded RNA is formed by base-pairing between complementary base sequence on the same strand.

RNA regulatory secondary structures have been detected in almost all living organisms ranging from viruses to Homo sapiens. The earlier discoveries of their regulatory role have been performed in model organisms such as the little worm $C.~elegans$ \cite{Lee1993} and in studies of the interaction between plants and viruses \cite{Ratcliff1997}.

Small noncoding RNA regulatory sequences are often characterized by the presence of hairpin structures. Hairpin structures have been investigated in quite different organims with a variety of methods. Examples of these studies can be found in Ref.s \cite{Schroth1995,Smith1995,Cox1997,Rice2000,lillo2002,Warburton2004,Leung2005,Chew2005,Spano2005,Lillo2007}

In this paper we detect candidate RNA secondary structures which are characterized by a minimal value of the free energy of the strucure in the folded state. In a first investigation the free energy is estimated by using Mfold, which is a reference software for the estimation of the free energy of RNA secondary structures. The free energy of the folded RNA structure is compared with the one numerically observed for a random RNA sequence obtained by shuffling the base pairs of the real one. This first investigation has been performed in a large set of 1212 complete genomes of viruses. 
We have chosen to perform a comprehensive investigate of these organisms because viruses present a variety of genomic structures and organization and there are indications that small RNA structures play important antiviral roles in plants and insects. Although the details about interactions between viruses and the host silencing RNA machinery remain poorly understood there is a mounting evidence that RNA motifs may play a crucial role in different aspects of the viral life-cycle. Results about the biological role of RNA secondary structures in different regions of different families of viruses are known only for a limited set of specific viruses with a focus on their coding regions
\cite{Simmonds1999,Tuplin2002,Thurner2004}. It is therefore useful to perform a  comprehensive investigation covering a large number of the complete genomes of viruses today available.

In a second investigation focused on the important viral family of herpesviridae we perform a search of RNA secondary structures by using RNAz \cite{Washietl2005}. This is a computer program based on thermodynamic and comparative genomic indicators. It has been used to detect evolutionary conserved RNA secondary structures in several organisms \cite{Washietl2005b}. We apply it to a large number of complete genomes of herpesviruses and search for candidate RNA secondary structures detected in coding and noncoding conserved genome regions. 

The paper is organized as follows: In Section 2 we illustrate the data set we investigate.  Section 3 discusses the method we use to search RNA secondary structures. In this section a detailed discussion is done about the limits of validity of random null hypothesis used to point out the predicted RNA secondary structures. Section 4 presents the results about the predicted RNA secondary structures detected in the investigated viruses when they are grouped according to their type of nucleic acid. Section 5 presents the investigation performed with the RNAz program in the family of herpesviruses and Section 6 briefly concludes.

%%%%%%%%
\section{The investigated data set}
\label{sec:1}
\begin{figure}
\resizebox{1\columnwidth}{!}{\includegraphics{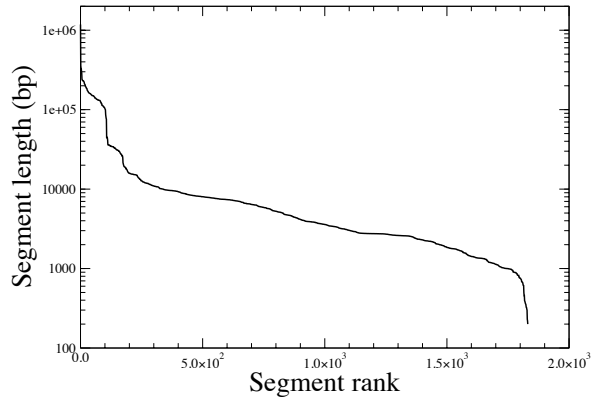}}
\caption{Rank plot of the length of the 1832 viral segments investigated in our study. The minimal and maximal length is 220 bp and 1.18 Mbp respectively.} 
\label{length}
\end{figure}

This study aims to investigate the presence of potential RNA secondary structures in a large set of complete genomes of viruses. For this reason, we analyze 1832 complete segments of 1212 complete viral genomes recently available. This set of complete genomes was downloaded from the GenBank database \cite{website}  in April 2006. Several classification systems of viruses exist being mainly based on phenotypic characteristics, including morphology, nucleic acid type, mode of replication, host organisms, and the type of disease they cause. In the present study we choose to focus on the Casjens and King \cite{Casjens1975} classification of viruses. They classified viruses into 4 groups essentially based on type of nucleic acid. Specifically they consider the following four groups: (i) double stranded DNA viruses, (ii) single stranded DNA viruses, (iii) double stranded RNA viruses, and (iv) single stranded RNA viruses. Viral complete genomes can be structured in one or more segments. Our set of completely sequenced segments of complete genomes of viruses comprises 310 dsDNA segments, 326 ssDNA segments, 896 ssRNA segments, and 300 dsRNA segments. The size of different segments of viral genomes ranges from 220 bp to 1.18 Mbp. The distribution of the segment length of investigated viruses is quite heterogeneous with many segments shorter than $10^4$ bp and a limited number of long segments. An overview of the length heterogeneity of considered segments is obtainable from Fig. \ref{length} where the ranking plot of the segment
lengths is shown. The total number of base pairs of our set of complete genomes is 28,56 Mbp. They include 23,92 Mbp for coding regions, and 4,65 Mbp for
noncoding regions. We point out this aspect because our analysis for each virus is performed by distinguishing the detected structures in coding and noncoding regions.

To complete the information about the analyzed set of viral segments, we provide a statistical summary of the CG content of viral segments. This information is provided in Table \ref{CG}. The data reported in the table shows that the average CG content observed in different groups and its standard deviation is not too markedly different in coding and noncoding regions of all considered groups.

% For tables use
\begin{table}
\caption{Descriptive statistics of the investigated viral segments according to the type of nucleic acid. Results are provided separately for coding (CDS) and noncoding (NCR) regions. The quantity $<CG>$ is the average value of the CG component of viral segments and $std\{CG\}$ its standard deviation. The symbol $\#$ indicates the number of viral segments while the length column indicates the total length in the considered group.}
\label{CG}       % Give a unique label
% For LaTeX tables use
\begin{tabular}{llllll}
\hline\noalign{\smallskip}
Region & group & $<CG>$ & std\{CG\} & \# & Length \\
\noalign{\smallskip}\hline\noalign{\smallskip}
CDS & all  & $0.447$ & $0.0658$ & $1780$ & $23.958$ Mbp\\
CDS & dsDNA  & $0.445$ & $0.0946$ & $303$ & $17.3$ Mbp\\
CDS & dsRNA  & $0.453$ & $0.0693$ & $286$ & $0.689$ Mbp\\
CDS & ssDNA  & $0.433$ & $0.0440$ & $323$ & $0.709$ Mbp\\
CDS & ssRNA  & $0.452$ & $0.0580$ & $868$ & $5.26$ Mbp\\
\noalign{\smallskip}\hline\noalign{\smallskip}
NCR & all  & $0.425$ & $0.0865$ & $1815$ & $4.65$ Mbp\\
NCR & dsDNA  & $0.396$ & $0.102$ & $304$ & $3.86$ Mbp\\
NCR & dsRNA  & $0.465$ & $0.0826$ & $300$ & $0.081$ Mbp\\
NCR & ssDNA  & $0.423$ & $0.0598$ & $326$ & $0.180$ Mbp\\
NCR & ssRNA  & $0.421$ & $0.0853$ & $885$ & $0.531$ Mbp\\
\noalign{\smallskip}\hline
\end{tabular}
% Or use
%\vspace*{5cm}  % with the correct table height
\end{table}

%\subsection{Subsection title}
%\label{sec:2}
%as required. Don't forget to give each section and subsection a unique label (see Sect.~\ref{sec:1}).

%%%%%%%%
\section{The search method}

We detect candidate RNA secondary structures in RNA sequences of the viral segments by computing the minimum free energy structures predicted by the Mfold
(version 3.2) software \cite{Zuker1989}. This widely used software estimates the difference between
the free energy of the unfolded state from the one of the folded state of a RNA sequence. Calculations are performed with the temperature parameter sets to 37 degree Celsius for all sequences.

The investigated segments of the complete genomes are scanned with Mfold by using a sliding window of 80 bp moving in steps of 40 bp. Each selected RNA sequence is folded as a linear RNA sub-sequence. For each obtained value of the free energy $\Delta G$ associated with each investigated 80 bp RNA sequence, we estimate a Z-score $Z=(\Delta G -<\Delta G_{shuf}>)/std\{\Delta G_{shuf}\}$ by comparing the observed free energy $\Delta G$ with the mean value $<\Delta G_{shuf}>$ and standard deviation $std\{\Delta G_{shuf}\}$ of the free energy  computed by performing 100 mononucleotide shuffling of the considered 80 bp RNA sequence.

The Mfold algorithm estimates the minimum free energy by adding a negative stacking energy of base pairs (which is stabilizing the secondary structure) and a positive energy term (which is destabilizing the secondary structure) associated with non-complementary bases \cite{Zuker1989}, i.e. hairpin loops, interior/bulge loops and multiloops. A stabilizing energy contribution comes form the stacking energy between two adjacent base pairs. This is the reason why Workman et al. \cite{Workman1999}, and Rivas et al. \cite{Rivas200}, have concluded that an appropriate null hypothesis as random RNA sequence should be generated by taking into account the empirically observed dinucleotide frequency.

\begin{figure}
\resizebox{1\columnwidth}{!}{\includegraphics{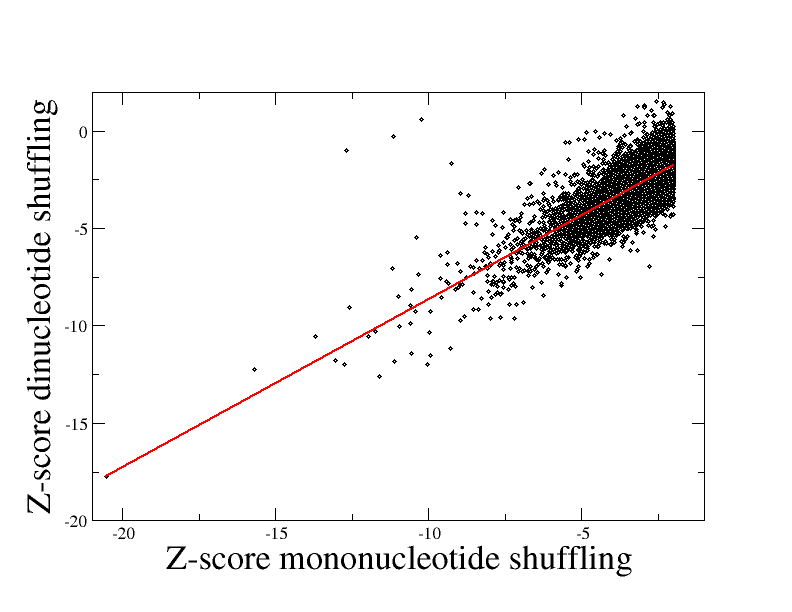}}
\caption{Z-score obtained with a null hypothesis based on dinucleotide shuffling as a function of the Z-score obtained with a null hypothesis based on a mononucleotide shuffling. Each circle represents the Z-score values obtained for a 80 bp sequence sampled from the set of the 378 largest viral segments under the condition that the mononucleotide Z-score is smaller than -2. The line is the linear regression of the set of points and it is characterized by a slope equals to 0.86. When we perform a linear regression on smaller sets of investigated points selected by the conditions $Z<-3$, $Z<-4$ and $Z<-5$ the value of the coefficient of the linear regression is 0.82, 0.77 and 0.73 respectively.} 
\label{Shuff}
\end{figure}

On the other hand, investigations of RNA secondary structures in the coding region of hepatitis C virus \cite{Tuplin2002} have shown that in members of the Flaviviridae mononucleotide shuffling and dinucleotide shuffling produce free energy values (and therefore Z-scores associated to them) which are remarkably similar. In other words, the use of the computationally more convenient mononucleotide shuffling is providing results compatible with the use of a dinucleotide shuffling in this family of viruses. Supported by this observation we have devised a test to assess whether a mononucleotide shuffling could be used as a good proxy for a dinucleotide shuffling in our investigations.
Specifically, we perform a dinucleotide shuffling for the set of 378 largest viral segments. The total length of these largest segments is 22.2 Mbp
which amounts to 84\% of the total length of the entire viral data set investigated by us. They have 18.9 Mbp of coding regions and 3.3 Mbp of noncoding regions. The dinucleotide shuffling is performed by using
the SHUFFLE program included in HMMER 2.2 package \cite{Durbin1998}. This program shuffles an
RNA sequence while preserving both mononuclotide and dinucleotide composition
exactly. It uses the Altschul and Erickson algorithm \cite{Altschul1985}. To limit computer time we estimate the relation between the Z-score obtained by considering a mononucleotide shuffling and the
Z-score obtained by considering a dinucleotide shuffling only for the sequences characterized by a mononucleotide Z-score smaller than -2. The results of our extensive test are shown in Fig. \ref{Shuff}. The figure clearly show that the mononucleotide shuffling is a good proxy of the dinucleotide shuffling in the large set of investigated viral segments. For this reason in the rest of this paper we will use the Z-score computed by using a mononucleotide shuffling of the investigated sequences.  

\section{RNA secondary structures in different groups of viruses}

For each RNA sequence of 80 bp we compute the optimal secondary structure according to the Mfold software and we associate to each RNA sequence a Z-score. The lower is the obtained Z-score the lower is the probability that the considered folding occurred by chance. By performing our large scale investigation, we systematically find a number of secondary structures characterized by a low value of mononucleotide Z-score.

% For tables use
\begin{table*}[t]
\caption{Number of HSs detected in coding and noncoding regions of all segments of investigated viruses as a function of the Z-score of each investigated 80 bp sequence. For comparison we also report the number of HSs detected in random sequences obtained by mononucleotide shuffling of the real ones. The HS-ratio is the ratio between the number of HSs detected in real data and the number of HSs detected in random sequences.}
\label{HS-All}       % Give a unique label
% For LaTeX tables use
\begin{tabular}{lllccccccc}
\hline\noalign{\smallskip}
source & region & group & $Z<-2$ & $Z<-2.5$ & $Z<-3$ & $Z<-3.5$ & $Z<-4$ & $Z<-4.5$ & $Z<-5$\\
\noalign{\smallskip}\hline\noalign{\smallskip}
real data & CDS  & all & $27514$ & $14726$ & $7604$ & $3805$ & $1923$ & $1015$ & $539$\\
mononucleotide shuffling & CDS  & all & $14461$ & $6424$ & $2709$ & $1094$ & $416$ & $160$ & $67$\\
HS-ratio real/shuffled data & CDS  & all & $1.90$ & $2.29$ & $2.81$ & $3.48$ & $4.62$ & $6.34$ & $8.0$\\
\noalign{\smallskip}\hline
real data & NCR  & all & $8721$ & $5563$ & $3676$ & $2373$ & $1615$ & $1076$ & $752$\\
shuffled data & NCR  & all & $2782$ & $1177$ & $515$ & $227$ & $97$ & $39$ & $12$\\
HS-ratio real/shuffled data & NCR  & all & $3.13$ & $4.73$ & $7.14$ & $10.4$ & $16.6$ & $28$ & $63$\\
\noalign{\smallskip}\hline
\end{tabular}
% Or use
\vspace*{1cm}  % with the correct table height
\end{table*}

To focus on a well defined part of the secondary structures, for each computed RNA structure of minimum free energy, we identify all hairpin structures (HSs) that are present in each structure. Here we present results on HSs which have a stem length ranging from 6 to 40 bp. This is done both for real segments and for the corresponding null hypothesis obtained by performing a mononucleotide shuffling of each 80 bp RNA sequence maintaining its nucleotide composition. 

We first analyze the number of HSs observed in real and random sequences by conditioning on the Z-score value associated with the folding of the RNA sequence. The number of HSs observed in RNA sequences characterized by a Z-score smaller then -2, -2.5, -3, -3.5, -4, -4.5 and -5, is given in Tables \ref{HS-All}. Results are summarized separately for coding and non-coding regions. Our data clearly indicate that the number of HSs detected in the real viral segments is much higher than the one observed for the corresponding null hypothesis obtained for random segments generated with mononucleotide shuffling of real data. A higher number than the one expected under the null hypothesis is observed both in coding and in non-coding regions. This feature is more pronounced when we condition the analysis to RNA sequences characterized by low values of the Z-score. In fact, the lower is the mononucleotide Z-score the larger is the HS-ratio of number of HSs observed in the native segments to the number of HSs observed in the random segments. This result is much more pronounced in noncoding than in coding regions. In fact when we condition the analysis to RNA sequences with Z-score smaller than -5 we observe a ratio of 8.0 in the coding regions whereas we observe a ratio of 63 in noncoding regions. In summary HSs are much more abundant in viruses than expected under a simple but representative null hypothesis. The abundance is quite remarkable in noncoding regions where the HS-ratio reaches a value as high as 63 when the analysis is conditioned to low values of the Z-score of the RNA sequence.

The results summarized in Table \ref{HS-All} for the complete group of viral segments can be analyzed in more detail in terms of the 4 groups we use to classify the investigated viruses. Several investigations of ssRNA viruses such as the hepatitis G virus \cite{Simmonds1999} or the hepatitis C virus \cite{Tuplin2002} and virus of the family Flaviviridae \cite{Thurner2004} have concluded that RNA secondary structures are present in the coding regions of these viruses and some of them are also evolutionary conserved. It is therefore of interest to check our set to discriminate whether the detected secondary structures which are present in coding regions are present in ssRNA viruses or if they are also present in different group of viruses.
% For tables use
\begin{table*}
\caption{HS-ratio for coding and noncoding regions of all viral segments conditioned on the Z-score value of the investigated sequence. We summarize the results also by grouping the viral segments according to their nucleic acid. The HS-ratio is not shown when the number of detected HSs is smaller than 10 in real or random data.}
\label{HS-groups}       % Give a unique label
% For LaTeX tables use
\begin{tabular}{lllccccccc}
\hline\noalign{\smallskip}
source & region & group & $Z<-2$ & $Z<-2.5$ & $Z<-3$ & $Z<-3.5$ & $Z<-4$ & $Z<-4.5$ & $Z<-5$\\
\noalign{\smallskip}\hline\noalign{\smallskip}
HS ratio & CDS  & all & $1.90$ & $2.29$ & $2.81$ & $3.48$ & $4.62$ & $6.34$ & $8.0$\\
\noalign{\smallskip}\hline\noalign{\smallskip}
HS ratio & CDS  & dsDNA & $1.60$ & $1.82$ & $2.09$ & $2.47$ & $3.02$ & $4.19$ & $4.9$\\
HS ratio & CDS  & dsRNA & $1.91$ & $2.26$ & $2.8$ & $4.6$ & $--$ & $--$ & $--$ \\
HS ratio & CDS  & ssDNA & $2.54$ & $3.23$ & $4.7$ & $5.3$ & $7.6$ & $--$ & $--$ \\
HS ratio & CDS  & ssRNA & $2.77$ & $3.73$ & $5.06$ & $6.49$ & $10$ & $14$ & $19$\\
\noalign{\smallskip}\hline
HS ratio & NCR  & all & $3.13$ & $4.73$ & $7.14$ & $10.4$ & $16.6$ & $28$ & $63$\\
\noalign{\smallskip}\hline\noalign{\smallskip}
HS ratio & NCR  & dsDNA & $2.83$ & $4.27$ & $6.22$ & $9.11$ & $16$ & $30$ & $58$\\
HS ratio & NCR  & dsRNA & $2.3$ & $2.3$ & $--$ & $--$ & $--$ & $--$ & $--$ \\
HS ratio & NCR  & ssDNA & $5.20$ & $9.3$ & $26$ & $--$  & $--$  & $--$ & $--$ \\
HS ratio & NCR  & ssRNA & $5.33$ & $7.78$ & $12$ & $12$ & $--$ & $--$ & $--$ \\
\noalign{\smallskip}\hline
\end{tabular}
% Or use
%\vspace*{5cm}  % with the correct table height
\end{table*}
In Table \ref{HS-groups} we summarize the HS-ratio for the 4 groups we use in our classification, namely dsDNA, dsRNA, ssDNA and ssRNA. The Table shows that the HS-ratio of ssDNA or ssRNA segments is higher than the one observed for dsDNA or dsRNA segments. In the Table we indicate with the symbol $--$ the case when the ratio is computed with a number of HSs detected in the real or shuffled segments smaller than 10. In other words these values have associated a large error in the estimation of the HS ratio and therefore we will not show them. The analysis of the Table shows a different behavior of the groups of viruses with respect to coding and noncoding regions. Specifically in noncoding regions the values of the HS-ratio are significantly higher than in coding regions. For example when $Z<-4$ in the coding regions of dsDNA viruses (the group where the best statistics is achieved due to the fact that these viruses are characterized by long genomes) we observe a HS-ratio equals to 3.02 whereas in noncoding regions the HS-ratio is equal to 16. A significant difference is also observed inside the same region (coding or non coding) when we distinguish among the different groups. For example when $Z<-3.5$ in coding regions dsDNA are characterized by a HS-ratio of 2.47 whereas ssRNA have a HS-ratio of 6.49. The other groups of dsRNA and ssDNA have intermediate values. Similarly in noncoding regions dsDNA are characterized by a HS-ratio of 9.11 whereas ssRNA have a HS-ratio of 12. Table \ref{HS-groups} shows that in coding regions of ssRNA and ssDNA viruses present a HS-ratio which is significantly higher than the one observed in dsDNA and dsRNA viruses. It is known in the literature that RNA secondary structures with known or potential biological role are present in the coding regions of ssRNA viruses \cite{Simmonds1999,Tuplin2002,Thurner2004}.
 
It is worth noting that the HS-ratio in dsDNA and dsRNA viruses is however significantly larger than one. This is an indication that the RNA secondary structures where these HSs are located might also have a potential biological role. Just to provide a comparative evidence that these values are significantly higher than one we have performed our analysis on one chromosome of a model organism. Specifically we have investigated chromosome V of the {\it C. elegans}. When we perform our analysis on the coding regions of this chromosome we obtain the following values for the HS ratio: 1.17 when $Z<-2$, 1.22 when $Z<-2.5$, 1.16 when $Z<-3$, 1.10 when $Z<-3.5$, 1.10 when $Z<-4$, 1.62 when $Z<-4.5$ and 1.86 when $Z<-5$. It is worth noting that noncoding regions of the chromosome V of the {\it C. elegans} also present high values of the HS-ratio. In fact when we perform our analysis on these regions we obtain 2.41 when $Z<-2$, 3.32 when $Z<-2.5$, 5.05 when $Z<-3$, 7.75 when $Z<-3.5$, 13.6 when $Z<-4$, 20.8 when $Z<-4.5$ and 38 when $Z<-5$. The analysis of the HS-ratio values obtained in coding and noncoding regions of the chromosome V of the {\it C. elegans} shows that the HS-ratio values obtained in coding regions are significantly smaller than the ones we have observed for the dsDNA and dsRNA viruses. This observation has motivated us to search for the presence of conserved RNA secondary structures in dsDNA viruses. 

\section{Conserved RNA secondary structures in the herpesvirus family}

The herpesviridae are a family of large encapsulated DNA viruses. Herpesvirus
genomes are circular or linear dsDNA up to approximately 250 kb in length
containing approximately between 70 and 220 genes. The herpesviridae are divided into three subfamilies: alpha, beta, and gamma herpesviruses \cite{McGeoch1995,McGeoch2000} according to their host range, cytopathology, and molecular phylogenetic analysis. All three groups have been found in primates including humans. 

We analyze here 21 completely sequenced genomes of herpesviruses and 1 large fragments. The herpesvirus genomes were downloaded from GenBank in July 2007 from the website \cite{website}. The set consists of 13 alpha herpesviruses, 3 beta herpesviruses and 6 gamma herpesviruses. The phylogenetic information about these viruses we use in our analysis is derived from McGeoch et al. \cite{McGeoch2006} and Davison et al. \cite{Davison2002}. In our analysis, we cluster herpesvirus genomes in 6 subgroups according to their genera (alpha-1, alpha-2, alpha-3, etc., for details see Table \ref{allign_herp8}). 

%TABELLA 4
\begin{table*}
\caption{6 herpesvirus groups of complete genomes belonging to different herpesvirus genera. Each distinct group is investigated with the RNAz program.}
\begin{center}
\begin{tabular}{l l l c c c c}
\hline
subfamily       &   generis   & species                        & CDS         & conserved    &   NCR      & conserved      \\ 
                &              &                                & length      & CDS          &   length   & NCR            \\
                &              &                                & (bp)        & ($\%$)       &   (bp)     & ($\%$)         \\

\hline
\hline
                &              &  Human herpesvirus 2           & 122543      &  93          &  32203     &  55          \\ 
\cline{3-7}
                &              &  Human herpesvirus 1           & 121248      &  94          &  31013     &  54          \\ 
\cline{3-7}
                & alpha 1      &  Cercopit. herpesvirus 1       & 119287      &  94          &  37502     &  47         \\ 
\cline{3-7}
                &              &  Cercopit. herpesvirus 16      & 118774      &  94          &  37713     &  46          \\ 
\cline{3-7}
                &              &  Cercopit. herpesvirus 2       & 117121      &  94          &  33594     &  47         \\ 
\cline{2-7}              
&  &  &              &             &              &                   \\ 
\cline{2-7}
                &              &   Equid herpesvirus 4          & 124329      & 64          &  21268      &  14        \\ 
\cline{3-7}                
                &              &   Equid herpesvirus 1          & 125445      & 63          &  24779      &  12        \\ 
\cline{3-7}
alpha           & alpha 2      &   Suid herpesvirus 1           & 108501      & 66          &  34960      &   8         \\ 
\cline{3-7}                
                &              &   Bovine herpesvirus 5         & 116183      & 66          &  21638      &  21        \\ 
\cline{3-7}                
                &              &   Bovine herpesvirus 1         & 114897      & 67          & 20404       & 16         \\ 
\cline{2-7}              
&  &  &              &             &              &                   \\ 
\cline{2-7}
                &              &  Gallid herpesvirus 3          & 134015      &  86        &    30255    &  30   \\
\cline{3-7}               
                & alpha 3      &  Gallid herpesvirus 2          & 137562      & 80         &    40312    &  40           \\
\cline{3-7}                
                &              &  Meleagrid herpesvirus 1       & 131798      & 89         &    27362    &  32   \\
\hline
&  &  &              &             &              &                   \\ 
\hline
                &              & Chimp. cytomegalovirus     &  192875       &  68   &  48212     & 28     \\ 
\cline{3-7} 
beta            & beta 1       & Human herpesvirus 5        &  186974       & 69      & 48671    & 27        \\ 
\cline{3-7}                 
                &              & Cercopit. herpesvirus 8    &  174220       &  66     &  47234    &  32           \\ 
\hline
&  &  &              &             &              &                   \\ 
\hline       
                &              &  Cercopit. herpesvirus 15   & 118815       &  73     &  52281      &  11          \\
\cline{3-7} 
                & gamma 1      &  Human herpesvirus 4        & 132121       &  67     &  39702      & 8                  \\
\cline{3-7}                
                &              &  Callitric. herpesvirus 3   & 104103       &  83     &  45593      &  12           \\ 
\cline{2-7}                 
gamma &  &  &              &             &              &                   \\ 
\cline{2-7} 
                &              & Porcine herpesviruses 1     & 64095       &  52      &  9105      &   3.5         \\ 
\cline{3-7} 
                & gamma 2     & Alcelap. herpesvirus 1     & 104327       & 32       & 26281      &   1.0           \\ 
\cline{3-7}
                &              & Equid herpesvirus 2        & 108637       & 31      &  75790      &   0.7           \\ 
\hline
\hline
%\hline
\end{tabular}
\end{center}
\label{allign_herp8}
\end{table*}

Specifically, we investigate 6 subgroups of genomes, each containing between 3 and 5 viral genomes. The analysis is performed as follows. We use a modified version of the Multiz-Tba package \cite{Blanchette2004} to generate a whole-genome alignment for each of the 8 groups of genomes. The Tba program uses a phylogenetic tree and the corresponding sequences as inputs, and outputs the resulting whole-genome alignment. High sequence conservation is not strictly needed for biological function of RNA secondary structures \cite{Bentwich2005,Pang2006} and therefore, we consider both medium and highly conserved genome regions in our analysis.

We use RNAz (version 1.0) \cite{Washietl2005} to detect consensus secondary structures for each of
the above described herpesvirus groups. Alignments are sliced in overlapping windows of size 120 and steps of 40 nt. Each series of windows starts at the
beginning of a TBA block. All the sequences with gap-content greater than 25\%
of gaps are discarded from the alignment before analysis. Furthermore, we discard all sequences with masked letters content greater than 10\%. This criterion is used for excluding repeat sequences marked by RepeatMasker \cite{Repeat}. RNAz is currently limited to analyze alignments up to six sequences. Finally, we use RNAz to analyze both the forward and the reverse complement sequences.

This kind of analysis might produce a certain number of false positive RNA secondary structure. We estimate the expected number of false positive by performing the following procedure. We use the program rnazRandomizeAln to shuffle the positions in an alignment. This shuffling removes any correlations arising from a native secondary structure and produces random alignment of the same length, the same base composition, sequence conservation, and gap patterns. The procedure is conservative providing a number of false positive higher than the one expected under a simpler random null hypothesis \cite{Washietl2005}.  
This procedure therefore gives us an estimate of the false-positive rate expected for a specific input alignment. The program tries to maintain local conservation pattern by shuffling only columns of the same degree of conservation, i.e. by shuffling the columns which show the same mean pairwise
identity. We therefore repeat the complete analysis with randomized alignment blocks. 

Sequence similarities between the species belonging to the same subgroup are mapped and used to determine the level of genome conservation between the viruses. A relatively high number of similar regions are conserved within genera and much lower number conserved among members of different genera. According to the sequence comparison method used, the alpha-1 and alpha-3 herpesvirus groups clearly share more coding regions (CDS) with detectable sequence homology than the other groups. We observe that the percentage of conserved CDS within alpha-1 herpesvirus generis is close to 94\% (see Table \ref{allign_herp8}). Note that the percentage reported in the table is computed on conserved region length without removing gaps or insertions. This percentage is ranging between 80\% and 89\% for alpha-3 herpesvirus group. The gamma-2 generis has the lowest number of conserved coding regions. Only approximately 40\% of the total coding regions are conserved. We also estimate the percentage of conserved noncoding regions between the genomes within genera. For example, within the gamma-2 group, approximately 2\% of the noncoding regions are conserved. The alpha-1 group contains the highest percentage of conserved noncoding regions.

The aim of our study is to select evolutionarily conserved motifs in subgroups of the Herpesviridae family. The threshold value we use for the "RNA class probabilities" is P=0.9 \cite{Washietl2005}. This is a rather stringent threshold value and it is useful to enhance the quality of the prediction. We observe that the number of conserved structures varies between herpesvirus subgroups. We find several conserved secondary structures in the coding regions of the gamma-1, beta-1A, alpha-1 and alpha-3 herpesvirus groups. In particular, in the coding region of the gamma-1 group we detect 29 conserved structures (see Table
\ref{P_GT_09_CDS}). In the coding regions of beta-1A group we detect 25 conserved structures and 19 and 15 in the coding regions of the alpha-1 and alpha-3 herpesvirus groups respectively. In coding regions of dsDNA viruses we therefore observe a number of conserved RNA secondary structures which have a potential biological role. The presence of evolutionary conserved RNA secondary structures is therefore not limited to ssRNA viruses.

%TABELLA 4.5
\begin{table*}
\caption{Rnaz hits with P$>$0.9 in coding regions of the 6 herpesvirus groups}
\begin{center}
\begin{tabular}{l l c c c c}
\hline
subfamily       &  generis    &                        &            &              &               \\
%\hline
                &               &         predicted      &   estimated false        &  Comparison         &   Comparison         \\
                &               &           structures   &     positives            &  with Rfam            &  with miRBase      \\
                &               &                        &                          &  database           &  database     \\
\hline
\hline
                & alpha 1      &   15      &    4     & 0          &  1    \\
\cline{2-6}   
alpha           & alpha 2      &    3      &   2         & 0           &  1              \\
\cline{2-6}   
                & alpha 3      &     19    &   5     & 0           &   0   \\
%\hline
%   &         &              &             &              &                   \\ 
%\hline
\hline
   beta               &  beta     & 25         &     9       &   0           &  2         \\
%\hline
%&  &  &              &             &                           \\ 
%\hline
\hline
gamma           & gamma 1     &      29      &   11         &        0      &    2       \\ 
\cline{2-6}   
                & gamma 2    &        12    &      0      &        0      &     0      \\ 
\hline
\hline
%\hline
\end{tabular}
\end{center}
\label{P_GT_09_CDS}
\end{table*}

We have extended our investigation on evolutionary conserved RNA secondary structures to the conserved non coding regions which are present in the 6 genera of investigated herpesviruses. Our investigation shows that, in this case, we observe only a small number of conserved structures in the noncoding regions. For example, the number of the conserved structures in the noncoding regions of the alpha-1 herpesvirus subgroup is only 3 (see Table \ref{P_GT_09_NCR}), in spite of the fact that this group has a relatively high
percentage of conserved NCRs.

%TABELLA 4.7
\begin{table*}
\caption{Rnaz hits with P$>$0.9 in noncoding regions of the 6 herpesvirus groups}
\begin{center}
\begin{tabular}{l l c c c c}
\hline
subfamily       &   generis    &                        &           &              &               \\
%\hline
                &               &         predicted      &   estimated false        &  Comparison         &   Comparison         \\
                &               &           structures   &     positives            &  with Rfam            &  with miRBase      \\
                &               &                        &                          &  database           &  database     \\
\hline
\hline
                & alpha 1      &   3      & 1        &       0        & 0          \\
\cline{2-6}   
alpha           & alpha 2      &    0      & 0        &        0       &  0            \\
\cline{2-6}     
                & alpha 3      &    8      &    0     &         0         &   0     \\
%\hline
%   &         &              &             &              &                   \\
\hline
 beta           &  beta 1     &    5      &      0     &      0        &        0   \\
%\hline
%&  &  &              &             &                           \\ 
\hline
gamma           & gamma 1     &     0       &      1      &       0       &     0      \\
\cline{2-6}   
                & gamma 2    &     0       &      0      &       0       &     0      \\ 
\hline
\hline
%\hline
\end{tabular}
\end{center}
\label{P_GT_09_NCR}
\end{table*}

To assess whether the detected conserved RNA secondary structures have a known biological role, we compare all the conserved RNA structures with the sequences of the Rfam Database \cite{Griffiths2005} and of the miRBase Database \cite{Griffiths2006}. 
The miRBase Database contains 4584 hairpin precursor miRNAs (miRBase release 9.2, May 2007), expressing approximately 4700
mature miRNA products, in primates, rodents, birds, fish, worms, flies, plants and viruses. The Rfam Database contains 32897 sequences of known ncRNAs (Rfam version 8.1, April 2007). We detect only 6 of the 4584 miRNA precursors in a global survey of the 6 herpesvirus genera. Our set of conserved RNA structures has no overlap with the structures annotated in the Rfam database.
Therefore, the large majority of the 103 RNA secondary structures detected in coding regions and of the 16 RNA secondary structures detected in noncoding regions have completely unknown biological role.

\section{Conclusion}
Our study shows that in viral genomes the hairpin structures of RNA secondary structures are more abundant than expected under a simple random null hypothesis. The random null hypothesis we use is admittedly simple but we have verified that it is a reliable proxy of the most accurate null hypothesis based on dinucleotide shuffling for the set of genomes we investigate. The detected hairpin structures of RNA secondary structures are present both in coding and in noncoding regions for the four groups of viruses we investigate in our comprehensive study. For all groups hairpin structures of RNA secondary structures are detected more frequently than expected for a random null hypothesis in noncoding rather than in coding regions. Differently from our previous results reported in \cite{Spano2005}, we observe that the detected hairpin structures are preferentially located in the noncoding regions. Therefore an approach based on combinatorial considerations such as the one of Ref. \cite{Spano2005} and a thermodynamically based approach like the present one lead to a different conclusion.

The amount of excess observed in coding regions is also of interest. This excess is in agreement with recent observation of the presence of evolutionary conserved RNA secondary structures detected in ssRNA viruses as the hepatitis G virus \cite{Simmonds1999}, the hepatitis C virus \cite{Tuplin2002} and viruses of the family Flaviviridae \cite{Thurner2004}. Our results indicate that this conclusion is not valid only for ssRNA viruses but rather a similar behavior should also be observed for other groups of viruses such as, for example, dsDNA viruses. To support this conclusion we have performed with the program RNAz a search of the evolutionary conserved RNA secondary structures in conserved coding and noncoding regions of a large set of complete genomes of herpesviruses. We detect a significant number of evolutionary conserved RNA sequences in conserved coding regions of herpesviruses while only a few structures are detected in conserved noncoding regions. In summary a large number of potential RNA secondary structures are predicted both in coding and noncoding regions of all groups of viruses with a clear preference for a location in noncoding regions. However, at least in herpesviruses, the degree of evolutionary conservation of these structure is more pronounced in coding than in noncoding regions.

\acknowledgments 
Authors acknowledge partial support from the NEST DYSONET 12911 EU project.

\end{document}